\documentclass[prd, onecolumn, nofootinbib, floatfix]{revtex4}

\usepackage{epsfig} 
\usepackage{amsmath}
\usepackage{amsbsy,color}

\newcommand{\beq}{\begin{equation}}
\newcommand{\eeq}{\end{equation}}
\newcommand{\beqa}{\begin{eqnarray}}
\newcommand{\eeqa}{\end{eqnarray}}

\newcommand{\ls}{\lesssim}

\def\ha{\frac{1}{2}}

\def \sn2{\left(S/N\right)^2}

\begin{document}

\title{The Synergy between Weak Lensing and Galaxy Redshift Surveys}
\author{Roland de Putter$^{1}$, Olivier Dor{\'e}$^{1}$, Masahiro Takada$^2$
\vspace{0.1cm}
}

\affiliation{$^{1}$Jet Propulsion Laboratory, California Institute of Technology, Pasadena, CA 91109\\
\& California Institute of Technology, Pasadena, CA 91125\\
$^{2}$Kavli Institute for the Physics and Mathematics of the Universe (Kavli IPMU, WPI), The University of Tokyo, Chiba 277-8582, Japan} 
\date{\today}

\begin{abstract}

We study the complementarity of weak lensing (WL) and spectroscopic galaxy
clustering (GC) surveys, by forecasting dark energy and modified gravity constraints
for three upcoming survey combinations: SuMIRe (Subaru Measurement of Images and Redshifts,
the combination of the
Hyper Suprime-Cam
lensing survey
and the Prime Focus Spectrograph redshift survey), EUCLID
and WFIRST.
From the WL surveys, we take into account both the shear and clustering
of the source galaxies and from the GC surveys, we use
the three-dimensional clustering of spectroscopic galaxies, including redshift space distortions.
A CMB prior is included in all cases.
Focusing on the large-scale, two-point function information,
we find strong synergy between the two probes.
The dark energy figure of merit from WL+GC
is up to a factor $\sim 2.5$ larger than from either probe alone.
Considering modified gravity, if the growth factor
$f(z)$ is treated as a free function, it is very poorly constrained by
WL or GC alone, but can be measured at the 
few percent level by the combination of the two.
On the other hand, for cosmological constraints
derived from (angular) power spectra and considering statistical
errors only,
it hardly matters whether the surveys overlap on the sky or not.
For instance, the dark energy figure of merit for overlapping surveys is at most $\sim 20 \%$ better than
in the disjoint case.
This modest gain can be traced to the fundamental fact that only a small fraction
of the total number of modes sampled by the GC survey (or by the WL survey) contributes to the
cross-correlations between WL and GC.

\end{abstract} 

\maketitle 

\section{Introduction}
\label{sec:intro}

Weak gravitational lensing surveys and spectroscopic galaxy redshift surveys are among the most promising
near-future
probes of dark energy, modified gravity and other cosmological physics.
Weak gravitational lensing, the subtle distortion of galaxy images by 
large-scale structure along the line of sight,
directly measures metric perturbations.
A measurement of cosmic shear, the large-scale correlations of the shear of galaxy images due to weak gravitational lensing,
therefore constrains cosmology through its sensitivity to the power spectrum of these metric perturbations
and through the dependence of the signal on the geometry of the universe.
Moreover, as a bonus, lensing surveys contain cosmological information
in the clustering of the lensing source galaxies.
Encouraging results have already been obtained from existing cosmic shear data, see
e.g.~\cite{Massetal07,Masseyetal07,Fuetal08,Schrabbetal09,huffetal11,kilbingeretal12,heymansetal13},
even though so far only a modest fraction of the sky (a few hundred square degrees) has been used
for these studies.

Spectroscopic galaxy surveys measure the three-dimensional matter distribution up to a
galaxy bias factor, which can be modeled on large scales, and are also sensitive to
the expansion history of the universe, as this determines the conversion from observed
angular positions and redshifts to three-dimensional coordinate positions.
Moreover, since a galaxy's redshift is determined not just by its cosmic distance,
but also by its peculiar velocity, the observed galaxy power spectrum or correlation function
receives a modification depending on the statistics of the large-scale velocity field.
These {\it redshift space distortions} make it possible to directly measure the growth rate of
large-scale structure.
As a cosmology probe, spectroscopic galaxy clustering surveys (see
e.g.~\cite{yorketal00,collessetal03,jonesetal09,parkinsonetal12,dawsonetal13,guzzoetal13}),
are already at a more mature level than weak lensing, with strong and robust cosmological constraints
so far coming especially from measurements of the baryon acoustic oscillations scale
\cite{eisetal05,coleetal05,beutleretal11,blakeetal11,andersonetal12}.

For both types of surveys, large steps forward will be made within the next decade,
thanks to a number of planned (or already ongoing) surveys that will vastly increase
the sky area and redshift range probed.
For weak lensing, these surveys include
the Kilo Degree Survey (KiDS, \cite{kids}), Dark Energy Survey (DES, \cite{des}),
the Subaru Hyper Suprime-Cam lensing survey (HSC, \cite{hsc}), LSST \cite{lsst} , EUCLID \cite{euclid} and WFIRST \cite{wfirst},
and for galaxy clustering, they include
HETDEX \cite{hetdex}, PFS \cite{pfsreport12}, DESI \cite{DESIwhitepaper}, EUCLID and WFIRST.

The imminent availability of high-quality galaxy clustering and weak lensing data
begs the question how much more can be learned about dark energy and modified gravity when
the two probes are combined. In this work, we will consider only the information
in the two-point statistics of the observed fields (shear, source galaxy density and spectroscopic
galaxy density) and restrict the analysis to large, quasi-linear scales.
However, we note that invaluable additional information may be encoded in
the data beyond the large-scale two-point function,
and that several promising methods for extracting this information
have been proposed, e.g.~\cite{yooseljak12,hikageetal12,hikageetal12b}.
Considering for simplicity two surveys covering an equal amount of sky (but not necessarily the same part),
one can study two distinct scenarios.

On the one hand,
if the survey areas are completely disjoint, one expects strong complementarity
for two reasons. First of all, somewhat trivially, the combination of surveys covers twice as much sky,
and therefore, heuristically, twice as many modes as any survey individually.
Secondly, and more importantly, the two probes have distinct sensitivities to
cosmological parameters so that combining them helps break degeneracies in
cosmological parameter space and allows for constraints that are much stronger than expected
solely based on the larger sky coverage (i.e.~uncertainties smaller than $\sigma/\sqrt{2}$ can be achieved, where $\sigma$ is the smallest
of the two individual probe uncertainties).

On the other hand, one can consider the scenario where the two surveys overlap fully. A potential downside
of this relative to the previous case is that this does not enlarge the total sky coverage
so that the probed volume is smaller than in the case of disjoint surveys, although we will show that
this is a small effect.
On the positive side, since both surveys now probe the same three-dimensional density modes, there is information
available in the cross-correlations between the two surveys. In the case of the the cross-correlations
between the shear modes and the spectroscopic galaxy density modes, this signal is equivalent to galaxy-galaxy lensing,
when the spectroscopic galaxies are at lower redshift than the lensing source galaxies.
By measuring the same density modes using different probes, one is effectively applying a multi-tracer
method \cite{mcdsel09}, and it should in principle be possible to extract certain cosmological information
without the limitation of sample variance.
While not the focus of this article, there are large additional advantages of overlapping surveys:
for instance, the imaging survey can provide target selection for the spectroscopic survey,
and the fact that the two surveys are subject to different systematics allows for
more robust measurements when they are combined,
(e.g., see \cite{hikageetal12,hikageetal12b} for such a method).
These same-sky benefits are not explicitly included in our analysis of statistical uncertainties only.

While it is clear that in both of the above cases, the combination of weak lensing and galaxy clustering
will improve cosmology constraints, two questions merit further study. (1) How strong exactly is this complementarity
for upcoming surveys (how large a boost in constraining power can be expected)?
and (2) how important is the overlap on the sky between the two surveys (i.e.~what is the difference in expected constraints
between the two scenarios discussed above)?
These questions have been studied to some extent in the literature
\cite{Guziketal:10,caibern12,gaztaetal12,kirketal13,fontetal13},
but especially the answer to question (2) varies between groups.
In this article, we study joint constraints from lensing and spectroscopic galaxy surveys
for three combinations of upcoming surveys:
Subaru Measurement of Images and Redshifts (SuMIRe), which is the combination of
the HSC lensing survey and the PFS spectroscopic galaxy survey
(both with the 8.2m Subaru telescope),
EUCLID \cite{euclid}, and WFIRST \cite{wfirst}. The latter two are satellite surveys
that will carry out both imaging and spectroscopic redshift programs.

Using the Fisher matrix formalism, we will focus our attention on forecasted constraints
on the dark energy equation of state, quantified by a Figure of Merit,
and on the growth factor of large-scale structure as a function of redshift, $f(z)$.
We will compare joint constraints to constraints from the weak lensing and spectroscopic survey
individually, always including a CMB prior from Planck,
and marginalizing over cosmological (and galaxy bias) parameters, including the sum of
neutrino masses.
While all three surveys will have full overlap between the imaging and spectroscopic components,
we will also study the hypothetical case of them being disjoint, in order to
quantify the importance (or lack thereof) of the cross-correlations (in other words, the same-sky benefits)
between the surveys.

The article is organized as follows. In section \ref{sec:method}, we will explain our forecast method, discussing the parameter space,
and paying specific attention to our approach for combining the information from angular power spectra and three-dimensional galaxy
power spectra. In section \ref{sec:surveys}, we briefly discuss the three survey combinations SuMIRe, EUCLID and WFIRST.
We present forecasted constraints for the different survey scenarios in
section \ref{sec:results}, and will explain these results in more detail in
section \ref{sec:why}. We conclude with discussion and a summary in section
\ref{sec:conclusions}.

\section{Method}
\label{sec:method}

We use the Fisher matrix formalism (see, e.g., \cite{TegTayHeav97}) to forecast cosmological constraints.
Our study takes into account data from two types of surveys.
On the one hand, we consider a sample of lensing source galaxies from
a weak lensing survey (WL). We assume these galaxies have photometric redshifts
(described by an unbiased Gaussian distribution with scatter $\sigma_z = \sigma_{z,0} (1 + z)$),
which are used to divide the sample into $N_{\rm tom}$ tomographic redshift bins,
defined by (photometric) redshift ranges $\{z_{{\rm tom}, i}^{\rm min},z_{{\rm tom}, i}^{\rm max}\}_{i=1}^{N_{\rm tom}}$.
In each bin, we then use two fields: the lensing shear as a function of position on the sky,
$\gamma$ (although in practice we capture its information in terms of the convergence $\kappa$),
and the relative overdensity of the number of source galaxies, $p$ ($p$ standing for photometric).
The clustering of the source galaxies is biased relative to the underlying dark matter field
so we model the galaxy bias as piecewise constant in redshift
and introduce a free galaxy bias parameter for each bin, $b^{(p)}_i$,
giving the value of the bias in the (true) redshift range $z_{{\rm tom}, i}^{\rm min} - z_{{\rm tom}, i}^{\rm max}$.
For simplicity, we assume this bias is constant in redshift
Since we consider only (quasi-)linear
scales, we take the bias to be scale-independent.

On the other hand, we consider data from a spectroscopic galaxy redshift survey (GC).
We divide this sample into $N_s$ redshift bins with bounds $\{z_{s, i}^{\rm min},z_{s, i}^{\rm max}\}_{i=1}^{N_s}$,and consider the galaxy overdensity
field, $s$, in each bin.
Since we assume the redshift in the spectroscopic sample to be measured with perfect
accuracy, we have full access to three-dimensional galaxy clustering information in each redshift slice.
As detailed below, to properly describe not only the information
in the 3D power spectrum of the spectroscopic survey, but also the cross-correlations
with the 2D $p$ and $\gamma$ fields, we split the Fisher matrix into two parts.
One part will be the usual Fisher matrix calculated for the 3D power spectrum in
redshift space (including redshift space distortions, Alcock-Paczynski effect, etc.),
while the other part will come from the angular auto- and cross-spectra with $\gamma$ and $p$,
using the projected spectroscopic galaxy density as a function of position on the sky
in each redshift bin.
To avoid double counting $s$-modes, we will remove the transverse modes already
included in the 2D Fisher matrix from the 3D Fisher matrix.
Just like for the photometric galaxy density, we introduce an independent, free galaxy bias
parameter, $b^{(s)}_i$, for each spectroscopic redshift bin (with label $i$).

To summarize, we consider three types of data: the shear $\gamma$ and photometric galaxy overdensity $p$
from an imaging/lensing survey, and the galaxy overdensity $s$ from a spectroscopic survey.
We will sometimes refer to the imaging survey and its data as WL (while it also contains galaxy clustering
information because of the $p$ field, the main goal of these surveys is the weak lensing shear signal)
and to the spectroscopic survey as GC.
As described in the following, we use cross- and auto-correlations between these fields (and between the
various redshift
slices) to construct a Fisher matrix and to forecast expected cosmological constraints.
In addition, all our results will have a Planck CMB prior included, where we neglect information
from CMB lensing because we want all the late-universe clustering information to come from the lensing
and galaxy clustering surveys of interest. We also neglect the correlation between CMB temperature anisotropies
and large-scale structure due to the Integrated Sachs-Wolfe effect. This is justified because the signal-to-noise
in this signal is low.

\subsection{2D Fisher Matrix}
\label{subsec:2d}

The 2D Fisher matrix encapsulates the information contained in
the angular auto- and cross-power spectra of the $p$, $\gamma$
and $s$ fields (or subsets thereof) in the different redshift bins.
We refer to the literature for the relevant equations describing the 2D Fisher matrix,
e.g.~\cite{caibern12}.
To calculate the angular power spectra, $C_l$, we employ the Limber approximation \cite{Limber:54,loveafsh08},
which expresses a spectrum as the line-of-sight integral over the product
of two kernels and the 3D matter power spectrum $P(k)$ (for the $p$ and $s$ fields, the kernel
contains a galaxy bias factor). We use the linear $P(k)$,
as obtained from CAMB \cite{LewChalLas00}, for all spectra except the
shear auto-spectra ($\gamma \gamma$, including cross-spectra between different
tomographic bins of course), for which we use the non-linear matter spectrum obtained
by applying the HaloFit \cite{HF03,taketal12} prescription to the linear power spectrum.
Using the non-linear signal for cosmic shear significantly increases the cosmological
information (see, e.g., \cite{phzpaper}) and is justified by the fact that shear directly measures
the underlying matter field and will therefore be easier to model in the mildly non-linear regime
than galaxy clustering, which would involve the additional complication of
non-linear galaxy bias.
We include $\gamma$ modes up to $\ell_{\rm max} = 2000$.
This is a common choice for cosmic shear forecasts,
although we note that modeling non-linear clustering and baryonic effects
to the required accuracy for this multiple range will be far from trivial.
For galaxy clustering, we choose a more
conservative cutoff, $\ell_{\rm max} = k_{\rm max} \, D(z_i)$, where $k_{\rm max} = 0.2 h/$Mpc
and $D(z_i)$ is the comoving angular diameter distance to the central redshift
of the $i$-th bin.
For both galaxy clustering and shear, we also apply a cutoff
$\ell_{\rm min} = 20$, because the largest angular scales
may be contaminated by systematics.
We will refer to the resulting Fisher matrix as ${\bf F}^{\rm 2D}_{\{\}}$,
where the curly brackets will contain the set of observables included, chosen from
$p, \gamma$ and $s$. All 2-point functions of the included fields will be used, both for the calculation
of the signal and that of the covariance matrix. This means that when a set of fields, e.g.~$\gamma$,
$p$ and $s$ are included in a single Fisher matrix, ${\bf F}^{\rm 2D}_{\{\gamma, p, s\}}$,
the fields are assumed to be measured on the same part of the sky and their covariances are
included. Fisher matrices for two non-overlapping surveys can be obtained by summing together two separate
Fisher matrices (see section \ref{subsec:combi}).

\subsection{3D Fisher Matrix}
\label{subsec:3d}

For the Fisher matrix of the 3D spectroscopic galaxy power spectrum in redshift space, $P_{ss}(k,\mu)$
($\mu$ being the cosine of the angle between the wave vector and the 
line-of-sight direction),
we follow the approach by \cite{seoeis07}.
The details of how we model $P(k,\mu)$ are as in \cite{pfsreport12},
except that we do not include the shot noise-like parameter $P_{\rm sn}$.
In particular, this means we apply an exponential damping in the Fisher matrix describing the effects
of non-linear clustering and redshift space distortions, and that we assume the use
of density field reconstruction \cite{eisetal07} to ameliorate this damping.
We include modes up to $k_{\rm max} = 0.2 h/$Mpc, but find our results to not be particularly sensitive
to small variations in $k_{\rm max}$ because the non-linear damping described above acts
as a {\it de facto} cutoff.
Finally, as mentioned above, we exclude $N_{\perp}$ transverse ($\mu \approx 0$) modes from our Fisher matrix,
where, for each bin in $k$ and $z$, $N_{\perp}$ is chosen to equal the number of $s$ modes
used in the 2D Fisher matrix. Specifically, this means that for a bin centered on $k_c$ and $z_i$,
we exclude a wedge $\Delta \mu = 2 \pi/(\Delta D(z_i) k_c)$ around $\mu = 0$, where
$\Delta D(z_i)$ is the width in comoving distance of the redshift slice centered at $z_i$. The same approach was followed in
\cite{caibern12,gaztaetal12}.
We will refer to the resulting Fisher matrix as ${\bf F}_{ss}^{\rm 3D*}$ (the star indicates that
transverse modes are left out).

\subsection{Combining Surveys}
\label{subsec:combi}

The main survey combinations we will consider are the following (the CMB prior is implicit):

\begin{itemize}

\item
{\it WL only}:  ${\bf F} = {\bf F}^{\rm 2D}_{\{\gamma,p\}} + {\bf F}^{\rm CMB}$\\
As explained above, the WL survey includes the information in the clustering
of the source galaxies, the $p$ field, in addition to lensing shear.
To clarify the above notation,
${\bf F}^{\rm 2D}_{\{\gamma,p\}}$ includes all cross- and auto-spectra of the types
$\gamma \gamma$, $\gamma p$ and $pp$.
We will also in some cases consider the case
where the information in the clustering of source galaxies is neglected and only $\gamma$ is used from the WL survey,
i.e.~replacing ${\bf F}^{\rm 2D}_{\{\gamma,p\}}$ by ${\bf F}^{\rm
     2D}_{\{\gamma\}}$. 

\item
{\it GC only}: ${\bf F} = {\bf F}^{\rm 3D*}_{ss} + {\bf F}^{\rm
  2D}_{\{s\}} + {\bf F}^{\rm CMB}$ \\
Alternatively, this matrix could be calculated by replacing ${\bf F}^{\rm 3D*}_{ss} + {\bf F}^{\rm 2D}_{\{s\}}$
by ${\bf F}^{\rm 3D}_{ss}$, the 3D galaxy power spectrum without transverse modes removed. We have calculated
this matrix as a consistency check, and find reasonable agreement between the two prescriptions,
lending support to our method of separating transverse modes and non-transverse ones.
Note that, while we refer to this case as {\it GC} for galaxy clustering, it refers to
the information in the spectroscopic survey only and thus does not include the additional galaxy clustering
information that would be available from the lensing source galaxies in an imaging survey.

\item
{\it GC + WL} (no overlap): ${\bf F} = {\bf F}^{\rm 3D*}_{ss} + {\bf F}^{\rm 2D}_{\{s\}} + {\bf F}^{\rm 2D}_{\{\gamma,p\}} + {\bf F}^{\rm CMB}$\\

\item
{\it GC + WL} (full overlap): ${\bf F} = {\bf F}^{\rm 3D*}_{ss} + {\bf F}^{\rm 2D}_{\{\gamma,p,s\}} + {\bf F}^{\rm CMB}$

\end{itemize}

\subsection{Parameters}
\label{subsec:params}

We consider a base set of $N_{\rm cosmo} = 9$ cosmological parameters,
$\{\omega_b, \omega_c, \Omega_\Lambda, \tau, \sigma_8, n_s, \Sigma m_{\nu}, w_0, w_a\}$
(the effect of the time-varying dark energy equation of state is implemented in CAMB
using the parametrized post-Friedmann formalism \cite{ppf}).
Note that this set includes the sum of neutrino masses, which is an unknown
that needs to be marginalized over. Its fiducial value is $\Sigma m_{\nu} = 0.15$ eV.
On top of these cosmological parameters, depending on which observables are taken into account,
we include the $N_{\rm tom}$ photometric galaxy bias, and the $N_s$
spectroscopic galaxy bias parameters.
Our dark energy constraints are calculated within this parameter space of a maximum of
$N_{\rm cosmo} + N_{\rm tom} + N_s$ parameters. We will summarize such constraints in terms
of the dark energy figure of merit (FOM, \cite{detf}),
\beq
{\rm FOM} = \left( {\rm Det}({\rm Cov}[w_0,w_a]) \right)^{-\ha}.
\eeq
Note that, unlike in \cite{detf}, we do not marginalize over spatial curvature, $\Omega_k$,
but do include $\Sigma m_{\nu}$.
As an aside, we find that generally the forecasted FOM looks significantly stronger
when $\Sigma m_{\nu}$ is fixed: typically a factor $2 - 3$ larger than
when it is properly marginalized over.

We will also study constraints on the linear growth rate of matter density perturbations,
\beq
f(z) \equiv \frac{d\ln D_m}{d\ln a} ,
\eeq
where $D_m$ is the linear growth factor of matter perturbations ($\delta_m(\vec{k},a) \propto D_m(a)$).
We study these constraints in the modified gravity (MG) scenario
where $f(z)$ is allowed to deviate from its value in  general relativity (GR),
$f(z) \approx \Omega_m(a)^\gamma$, with $\gamma \approx 0.55 + 0.05 (1 + w_0 + \ha w_a)$,
\cite{linder05}.
Given an amplitude $\sigma_8$ of the linear power spectrum at $z=0$, the amplitude of perturbations
at $z>0$ ($\sigma_8(z)$) is computed based on our choice of $f(z)$. 
In
other words,
we force the amplitude of perturbations to be consistent with our growth factor parametrization.
Moreover, for a given $\sigma_8$, the modified growth is taken into account to calculate the correct,
corresponding primordial amplitude of perturbations, $A_s$. This way, our treatment of CMB data
is also consistent with $f(z)$.

To parametrize $f(z)$ in the modified growth case, we introduce $N_f = N_s + 2$ parameters, $\{f_i\}_{i = 0}^{N_s+1}$,
describing a piecewise constant deviation from the GR value.
We thus have one free parameter for each spectroscopic redshift bin,
i.e.~$f_i$ describes the growth factor in the redshift range $z = z_{s,i}^{\rm max} - z_{s,i}^{\rm max}$
for $i = 1, ..., N_s$, and, in addition, $f_0$ is the growth factor in the range $z = 0 - z_{s,1}^{\rm min}$
and $f_{N_s + 1}$ the growth factor for $z > z_{s,N_s}^{\rm max}$ (we assume that the growth history returns
to that of GR well before $z \approx 1100$ so that the primary CMB
anisotropies are not affected).
The inclusion of free growth at redshifts beyond the range probed by galaxy clustering
will have strong implications for the ability of WL or GC individually to constrain the low redshift
growth history because the amplitude of matter fluctuations at the largest redshift probed by the spectroscopic
survey is now no longer determined by the CMB measurement ($f_{N_s+1}$ affects the translation of
the amplitude of perturbations at CMB last scattering to that at low redshift).
See, e.g., \cite{hojjetal12} for a discussion of the effects of marginalizing
over growth at high redshift.
While we do not include this information in the present study, we do note that
CMB lensing may help constrain $f_{N_s+1}$, at least to some degree.
When presenting our MG constraints on the growth factor,
we marginalize over $\Sigma m_{\nu}$, $w_0, w_a$ (in modified gravity, the {\it effective} dark
energy equation of state should be considered
a parametrization of the expansion history) and the other parameters, so that a maximum
of $N_{\rm cosmo} + N_{\rm tom} + N_s + N_f$ parameters are included in the Fisher matrix.
We note that, if instead we were to fix $\Sigma m_\nu$ when constraining growth, the constraints would be
stronger, but only by $\lesssim 30 \%$ (on the other hand, marginalizing over growth does strongly degrade
the neutrino mass constraint).

\section{Surveys}
\label{sec:surveys}

We make predictions for the following three combinations of WL and GC surveys.

\subsection{SuMIRe}
\label{subsec:sumire}

The Subaru Measurement of Images and Redshifts (SuMIRe) combines the weak lensing/imaging
data from the Hyper Suprime-Cam (HSC) survey and the spectroscopic data from
the Prime Focus Spectrograph (PFS) cosmology survey. The survey specifications (and fiducial galaxy bias) we use
can be found in our previous publication,
\cite{phzpaper}. We choose the following binning,
\begin{itemize}
\item
$N_{\rm tom} = 3$: $z_{{\rm tom},i}^{\rm min} - z_{{\rm tom},i}^{\rm max} = 0 - 0.6, \, 0.6 - 1, \, 1 - 4$
\item
$N_s = 7$: \, \,$z_{s,i}^{\rm min} - z_{s,i}^{\rm max} = 0.6 - 0.8, \, 0.8-1, \, 1-1.2, \, 1.2-1.4, \, 1.4-1.6, \, 1.6-2, \, 2- 2.4$
\end{itemize}

\subsection{EUCLID}
\label{subsec:euclid}

The EUCLID satellite mission will provide both weak lensing and spectroscopic galaxy clustering
data. 
Note that the EUCLID spectroscopic survey will be done using slit-less
low-resolution spectroscopy,
which does not need a pre-imaging survey to find the targets. 
We again follow the specifications outlined in \cite{phzpaper}. The binning choice is,

\begin{itemize}
\item
$N_{\rm tom} = 6$: $z_{{\rm tom},i}^{\rm min} - z_{{\rm tom},i}^{\rm max} = 0 - 0.4, \, 0.4 - 0.8, \, 0.8 - 1.2, \, 1.2 - 1.6, \, 1.6 - 2, \, 2 - 3.5$
\item
$N_s = 14$: \, \,$z_{s,i}^{\rm min} - z_{s,i}^{\rm max} = 0.65 - 0.75, \, 0.75 - 0.85, \, 0.85- 0.95,\, 0.95- 1.05,\, 1.05- 1.15,\, 1.15- 1.25,\, 1.25- 1.35,\, 1.35- 1.45,\, 1.45- 1.55,\, 1.55- 1.65,\, 1.65- 1.75,\, 1.75- 1.85, \, 1.85 - 1.95, \, 1.95 - 2.05$
\end{itemize}

\subsection{WFIRST}
\label{subsec:wfirst}

Finally, we consider the WFIRST satellite mission, which will also provide both weak lensing and galaxy clustering
information. The WFIRST spectroscopic survey will use slit-less spectroscopy.
Our assumed survey specifications mostly follow \cite{wfirst}.
Specifically, we assume both the lensing survey and the redshift survey will cover $2000$ deg$^2$.
For the lensing survey, we assume an angular number density $\bar{n}_A = 70 \, {\rm arcmin}^{-2}$,
and the same redshift distribution as we assumed for HSC ($\langle z \rangle = 1$). We assume the source
galaxies have galaxy bias $b^{(p)}(z) = 1$ and photometric redshift scatter
$\sigma_z(z) = 0.04 (1 + z)$ (compared to $0.05 (1 + z)$ for the previous two surveys).
For the spectroscopic sample,, we assume a fiducial galaxy bias $b^{(s)}(z) = \sqrt{1 + z}$.
We use the spectroscopic galaxy number density specified in Table 2-2 of \cite{wfirst}.
Finally, the binning choices are,
\begin{itemize}
\item
$N_{\rm tom} = 6$: $z_{{\rm tom},i}^{\rm min} - z_{{\rm tom},i}^{\rm max} = 0 - 0.4, \, 0.4 - 0.8, \, 0.8 - 1.2, \, 1.2 - 1.6, \, 1.6 - 2, \, 2 - 3.5$
\item
$N_s = 18$: \, \,$z_{s,i}^{\rm min} - z_{s,i}^{\rm max} = 1.075-1.175, \, 1.175-1.275, \, 1.275-1.375,\, 1.375-1.475,\, 1.475-1.575,\, 1.575-1.675,\, 1.675-1.775,\, 1.775-1.875,\, 1.875-1.975,\, 1.975-2.05,\, 2.05-2.15,\, 2.15-2.25, \, 2.25-2.35, \, 2.35-2.45, \, 2.45-2.55, \, 2.55-2.65, \, 2.65-2.75, \, 2.75-2.85$
\end{itemize}

\section{Results}
\label{sec:results}

\subsection{SuMIRe}

\subsubsection{Dark Energy}
\label{subsubsec:de}

\begin{figure*}
  \begin{center}{
  \includegraphics[width=0.48\columnwidth]{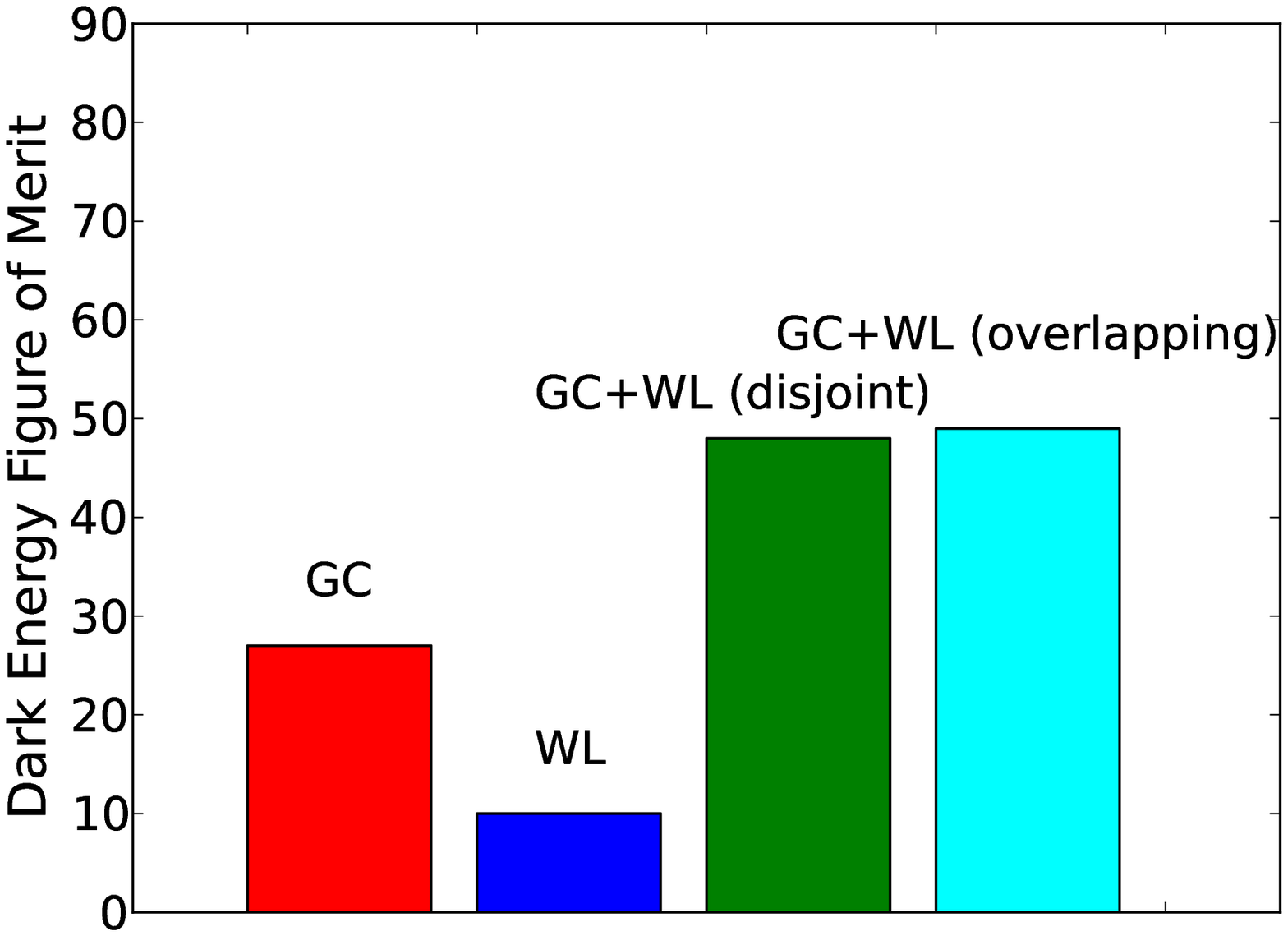}
  \includegraphics[width=0.48\columnwidth]{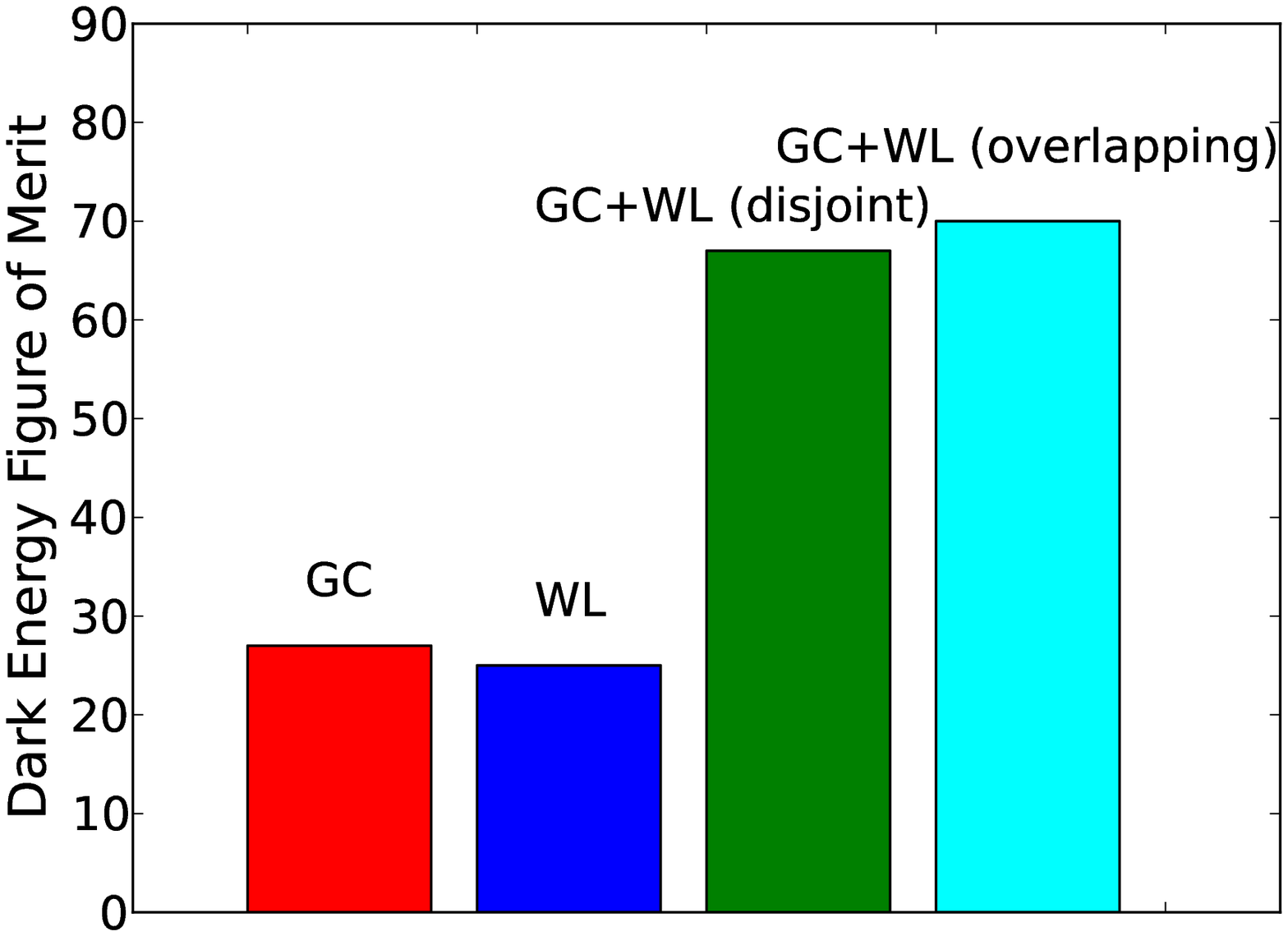}
  }
  \end{center}
  \caption{Forecasted SuMIRe dark energy figure of merit (see text for definition), marginalized over neutrino mass.
  Within each figure, from left to right, constraints are shown for
  GC: the spectroscopic survey only (PFS), WL: the lensing survey only (HSC),
  GC+WL(disjoint): the combination assuming they do not overlap on the sky,
  and GC+WL(overlapping): the combination when they do overlap (as they will in reality).
  All constraints include a Planck CMB prior.
  Combining the weak lensing and galaxy clustering surveys strongly improves the dark energy constraint.
  However, overlap between the surveys contributes negligibly.
  {\it Left panel:}
  WL includes shear only. {\it Right panel:} WL also includes the clustering of the lensing
  source galaxies, based on their photometric redshifts.}
  \label{fig:sumire fom}
\end{figure*}

We consider first the dark energy figure of merit for the different survey combinations
possible with HSC and PFS (SuMIRe). Figure \ref{fig:sumire fom}
shows constraints for GC (PFS) only, WL (HSC) only and for GC + WL, with and without overlap
(of course the actual surveys will overlap). The left panel shows the case where the clustering of
lensing source galaxies is not included on the WL side. In this case, while both surveys
individually (an implicit CMB prior is always included) deliver strong dark energy constraints,
GC has significantly more constraining power. The two bars on the right (of the left panel)
show that substantial improvements can be achieved by combining the two surveys, almost doubling the
FOM compared to the case of GC alone. However, very little of this complementarity comes from
the cross-correlations between the two sets of observables, and, accordingly,
the difference in FOM between the overlapping and non-overlapping scenarios is not noteworthy.
We will address the reasons why this is the case in Section \ref{sec:why}.

The right panel shows the case where all information from the imaging survey is used, i.e.~both
the shear and the clustering of source galaxies. With this included,
WL alone is competitive with GC alone. There is again strong complementarity when the two probes are added,
but the overlap still does not matter much.

The above of course does not take into account other benefits of the overlap between the two surveys.
For example, in the case of SuMIRe, HSC imaging will provide 
an ideal multi-color catalog of galaxies to find targets for the
follow-up spectroscopic PFS survey,
which
is extremely important. Moreover, if the photometric redshifts of the imaging survey
cannot be properly calibrated using a deep spectroscopic training/validation sample,
the cross-correlations with the GC survey can be used to improve the photo-$z$
calibration and thus make WL a stronger probe of dark energy, see, e.g.,
\cite{newman08,mathnewman10,schulz10,mattnewman12,mcwhite13,phzpaper}.

\subsubsection{Growth Factor}
\label{subsubsec:fg}

We next turn to constraints on the growth history, $f(z)$, in a modified gravity scenario,
considering the bounds on the growth parameter in each spectroscopic redshift bin.
As discussed previously, we marginalize over the growth factor at redshift both below and above the redshift
range where galaxies are observed by PFS. This has important implications. 
In particular, marginalization over the growth parameter, $f_{N_s+1}$,
at $z > z_{s,N_s}^{\rm max} = 2.4$ implies that,
even though the CMB measures both cosmological parameters and
the amplitude of perturbations in the early universe,
the CMB does not determine the amplitude of perturbations at redshifts where we observe
large-scale structure (GC or WL). Schematically,
GC measures the combinations $b^{(s)}_i \sigma_{8,i}$ and $f_i \sigma_{8,i}$
(where $i$ labels the redshift bin).
With $\sigma_{8,i}$ now unknown despite the CMB prior,
both galaxy bias and the growth factor
remain unconstrained.
Since WL, through its dependence on the amplitude of matter perturbations,
$\sigma_{8,i}$, is only sensitive to a degenerate combination
of growth factor parameters (i.e.~an integral over redshift of $f(z)$),
lensing alone (+CMB) yields rather poor growth factor constraints whether
$f_{N_s + 1}$ is left free or not, although constraints are better with
this parameter fixed.

\begin{figure*}
  \begin{center}{
  \includegraphics[width=0.48\columnwidth]{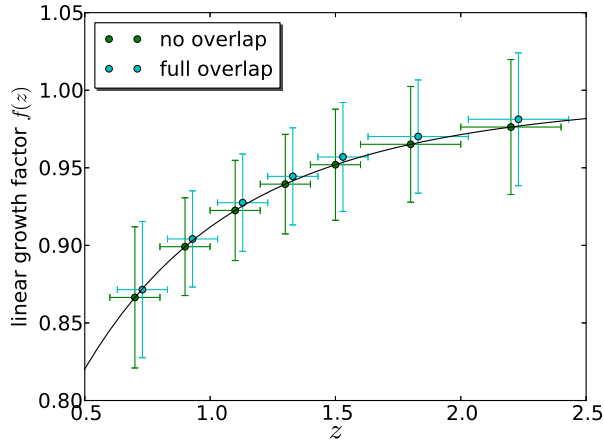}
  }
  \end{center}
  \caption{Forecasted constraints from SuMIRe WL + GC on the growth factor $f(z) = d\ln D/d\ln a$,
  modeling it as a free, piecewise constant function (i.e.~allowing for growth different than
  in general relativity), and marginalizing over neutrino mass and the effective equation
  of state parameters $w_0, w_a$ (see text). A Planck CMB prior is included.
  Projected uncertainties are indicated by the vertical bars, while the horizontal bars indicate the
  bin widths.
  Constraints from the spectroscopic galaxy clustering survey alone (PFS) or from the weak
  lensing survey alone (HSC) are not shown because they are very weak, typically $\sigma(f(z)) \gg 1$ (since we also allow freedom in $f(z)$ at high redshift),
  but combining them yields uncertainties at the $4$ percent level. As shown, whether or not the surveys overlap on the sky
  barely impacts the forecasted constraints.}
  \label{fig:sumire fgro}
\end{figure*}

When the two probes are combined, the degeneracy discussed above is broken, and very strong ($\gtrsim 3 \%$)
constraints can be obtained in all bins.
Figure \ref{fig:sumire fgro} shows these constraints both for the case of overlapping
and disjoint surveys. The horizontal bars indicate the bin widths.
While these constraints are strong, and will provide a strong test of general relativity,
it again does not matter whether the surveys overlap or not.

\subsection{EUCLID}

\subsubsection{Dark Energy}
\label{subsubsec:de euclid}

\begin{figure*}
  \begin{center}{
  \includegraphics[width=0.48\columnwidth]{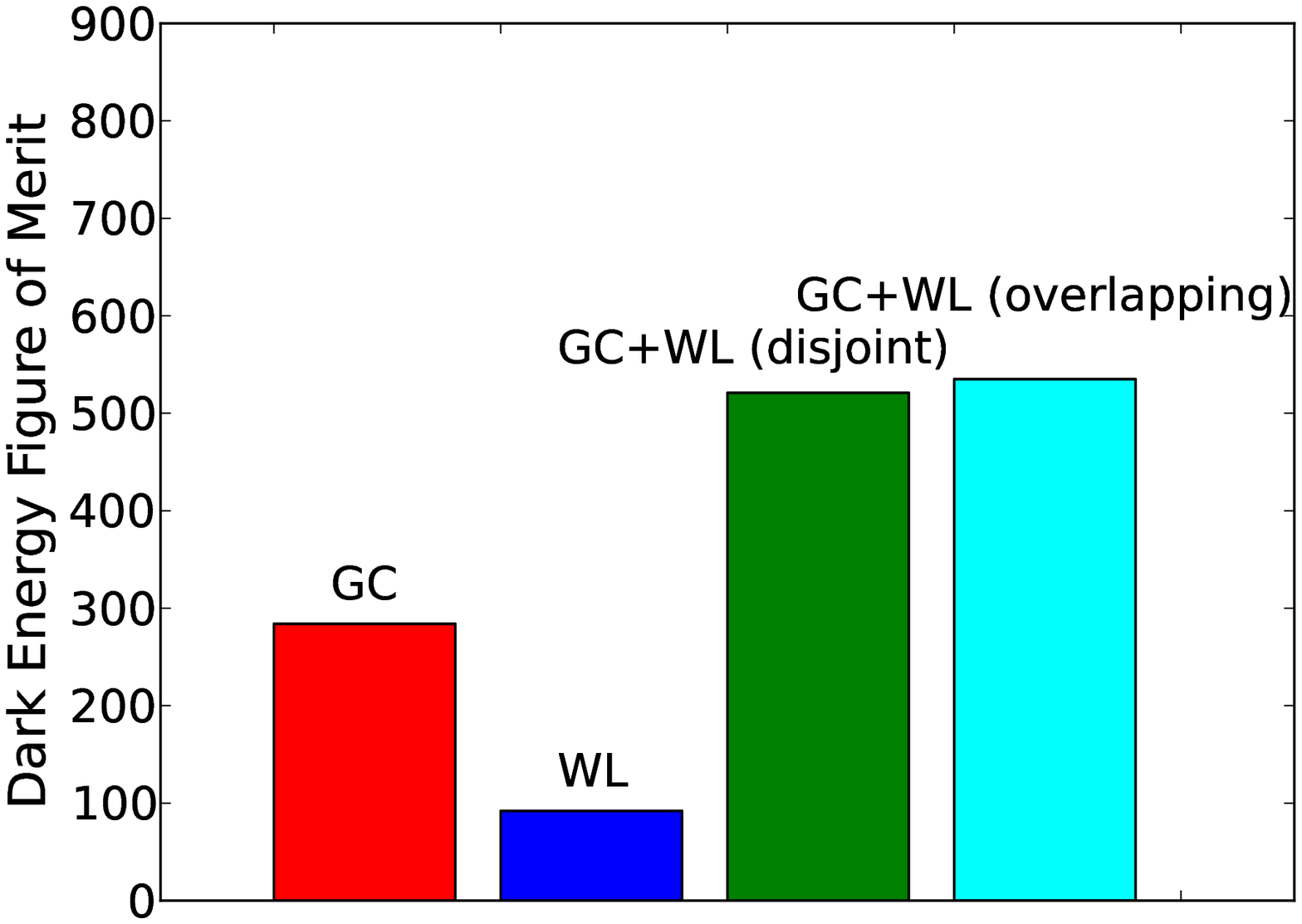}
  \includegraphics[width=0.48\columnwidth]{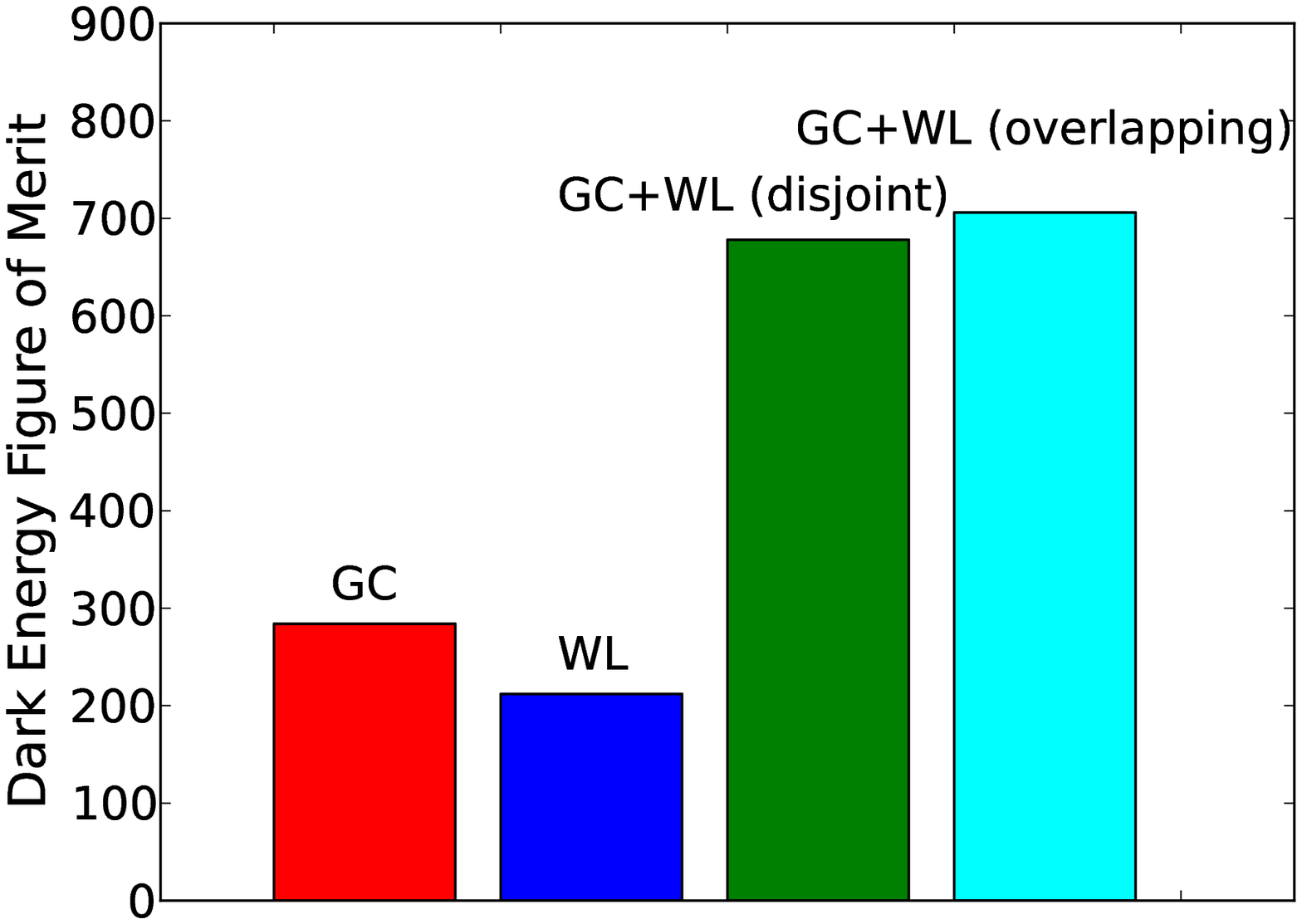}
  }
  \end{center}
  \caption{Dark energy figure of merit. Same as Figure \ref{fig:sumire fom}, but for the EUCLID
  weak lensing and galaxy clustering surveys. As for SuMIRe, strong complementarity between the two probes
  is found, while the overlap
  between surveys does not play a significant role.
  {\it Left panel:}
  WL means shear only. {\it Right panel:} WL also includes the clustering of the lensing
  source galaxies, based on their photometric redshifts.}
  \label{fig:euclid fom}
\end{figure*}

Figure \ref{fig:euclid fom} shows the forecasted dark energy constraints
for EUCLID (cf.~Figure \ref{fig:sumire fom}).
Focusing on the right panel, where the WL part of the survey includes
information from the clustering of lensing source galaxies (see the left panel
for the case where this information is neglected),
we see that WL and GC surveys on their own give comparable constraints,
as was the case for SuMIRe.
We do wish to note, however, that the comparison between the two is strongly dependent
on the treatment of non-linear scales for each probe. As a reminder, for GC,
we apply a non-linear cutoff, ensuring that little information is included
from the non-linear regime. WL shear, on the other hand, includes modes up to
$\ell_{\rm max} = 2000$, and uses the information in the (HaloFit) non-linear matter power spectrum.
It thus probes rather far into the non-linear regime, unlike GC. In both cases, we have tried to follow
the more or less standard choices adopted in the literature, to ease comparison.

Considering the synergy between the weak lensing and galaxy clustering components, the picture
is qualitatively the same as for SuMIRe: combining WL with GC strongly
improves dark energy constraints compared to the individual surveys,
but little is gained from overlap in sky coverage.

\subsubsection{Growth Factor}
\label{subsubsec:fg euclid}

\begin{figure*}
  \begin{center}{
  \includegraphics[width=0.48\columnwidth]{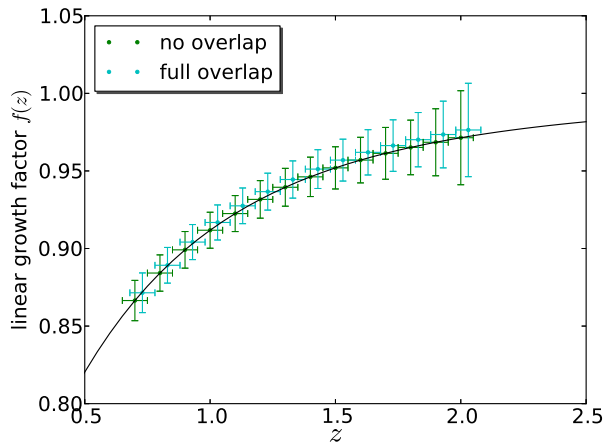}
  }
  \end{center}
  \caption{Projected growth factor constraints. As Figure \ref{fig:sumire fgro},
  but for the EUCLID
  weak lensing and galaxy clustering surveys. Only the combination of the two probes
  allows for strong growth factor constraints (forecasts for individual probes not shown). The constraints are effectively independent of
  whether or not the surveys overlap on the sky.}
  \label{fig:euclid fgro}
\end{figure*}

The growth factor constraints are shown in Figure \ref{fig:euclid fgro}
(cf.~Figure \ref{fig:sumire fgro}).
While constraints from any probe individually are again poor ($\sigma(f(z)) \gg 1$),
combining them, the growth factor can be measured to $1 \% - 2 \%$
in each spectroscopic galaxy bin. Whether or not the surveys overlap
again has little relevance.

\subsection{WFIRST}
\label{subsec:wfirst}

\begin{figure*}
  \begin{center}{
  \includegraphics[width=0.48\columnwidth]{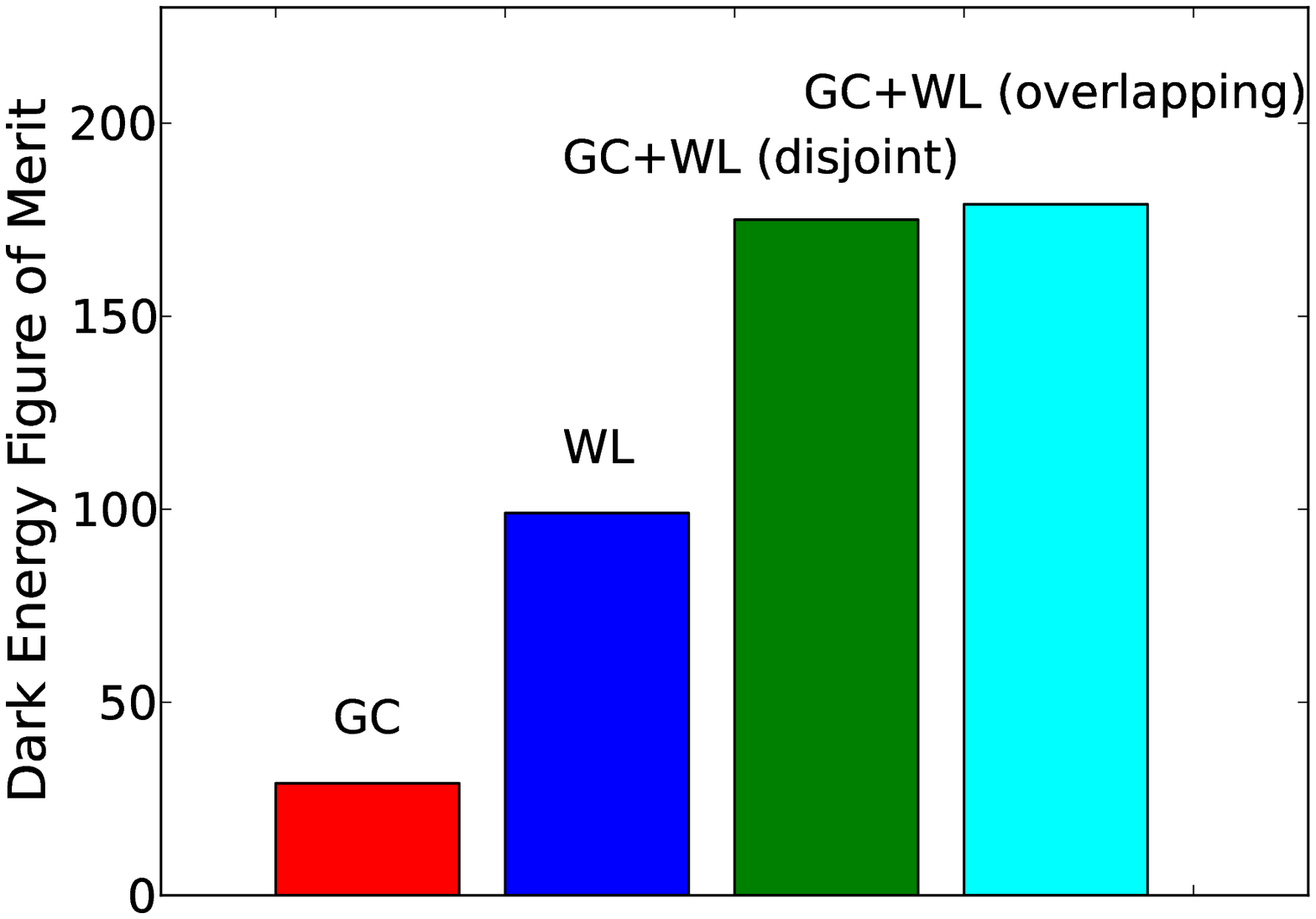}
  \includegraphics[width=0.48\columnwidth]{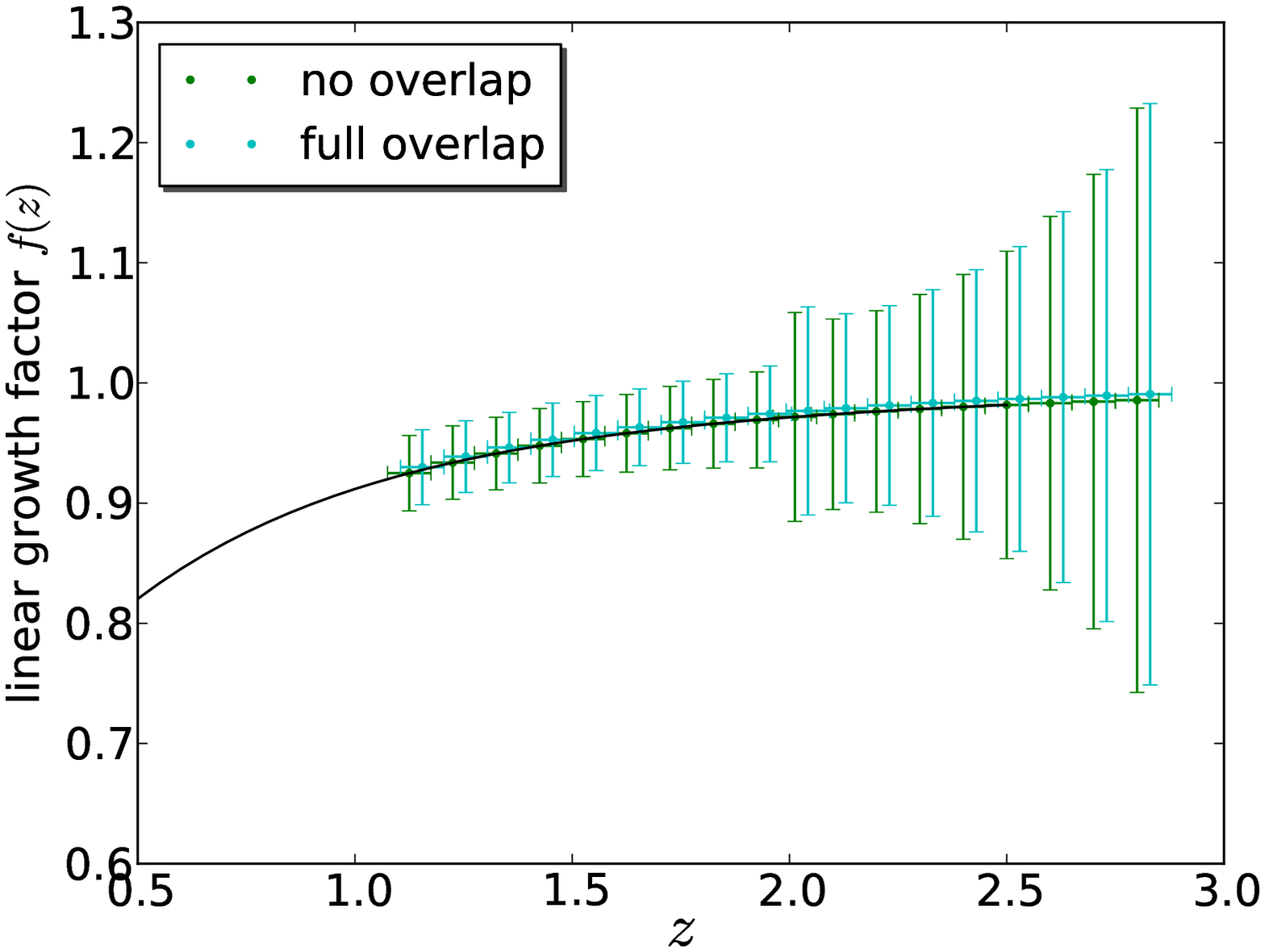}
  }
  \end{center}
  \caption{{\it Left panel:} Forecasted WFIRST dark energy figure of merit. WL here includes
  the clustering of the lensing source galaxies, cf.~right panels of
  Figures \ref{fig:sumire fom} and \ref{fig:euclid fom}.
   {\it Right panel:} WFIRST joint WL + GC growth factor constraints (WL only and GC only
   constraints on $f(z)$ are not shown as they are very poor, $\sigma(f(z)) \gg 1$), cf.~Figures \ref{fig:sumire fgro}
   and \ref{fig:euclid fgro}.}
  \label{fig:wfirst results}
\end{figure*}

Finally, we show both the dark energy and growth factor results for WFIRST in Figure \ref{fig:wfirst results}.
In all cases, the lensing survey is assumed to use both the shear information and the clustering of source galaxies.
Comparing the dark energy results (left panel) to SuMIRe (which has similar sky coverage), we find that the
WFIRST spectroscopic survey looks comparable to PFS (SuMIRe).
Comparing the results from the imaging survey component of WFIRST and SuMIRe, on the other hand,
we find that the WFIRST lensing survey is significantly stronger than HSC. This is mainly
explained by the large number density of the WFIRST source galaxies, $\bar{n}_A = 70 \, {\rm arcmin}^{-2}$
vs.~$\bar{n}_A = 20 \, {\rm arcmin}^{-2}$ for HSC. In addition, we have assumed slightly better photometric
redshifts for WFIRST ($\sigma_z(z) = 0.04 (1 + z)$ vs. $\sigma_z(z) = 0.05 (1 + z)$ for SuMIRe and EUCLID).
The joint dark energy constraints from WFIRST are much stronger than for SuMIRe, mainly because of the much more
powerful imaging survey.

Considering next the growth factor constraints (right panel, Figure \ref{fig:wfirst results}),
the WL and GC individually are again unable to place meaningful constraints (and are therefore not shown in the figure), but
for the combination of the two,
we find relative uncertainties in the range $3 \% - 25 \%$ (from low to high redshift).
The joint constraints are thus comparable to those from SuMIRe at the low redshift end, but significantly weaker
towards the highest redshifts.
To explain this, we first note that
the $f(z)$ bounds are mainly driven
by the spectroscopic survey, while the main role of the lensing survey is to help break degeneracies between $\sigma_8(z), b^{(s)}(z)$ and $f(z)$.
While both spectroscopic
surveys cover a deep, and equally broad redshift range ($z = 0.6 - 2.4$ for PFS and $z = 1.075 - 2.85$ for WFIRST),
the number density of galaxies in WFIRST drops
to $\bar{n} \lesssim 1 \times 10^{-4} (h^{-1} {\rm Mpc})^{-3}$ at $z > 2$, while the PFS galaxy density is
$\bar{n} > 3 \times 10^{-4} (h^{-1} {\rm Mpc})^{-3}$ at all redshifts.
This mostly explains the degradation of constraints at the high redshift end of WFIRST.
However, at all redshift, the difference in redshift bin width between the two surveys
also has the effect of making WFIRST appear weaker.
For SuMIRe, we used bins with width $\Delta z = 0.2$, while here we use $\Delta z = 0.1$, thus halving the volume available
per bin and weakening the constraints within bins (but giving a larger number of independent $f(z)$ measurements).
If we had used equal bin widths, the WFIRST $f(z)$ constraints would thus be stronger than those from SuMIRe at low
redshift. At the high redshift end, the number density is the dominant effect however.

In conclusion, considering both the dark energy and growth factor constraints, we again find strong complementarity between
weak lensing (plus source clustering) and spectroscopic galaxy clustering. However, as for the other surveys,
the role of the overlap between surveys is very limited.

\subsection{Uncertainty in Photometric Redshift Distribution}
\label{subsec:phz}

We have seen in the previous subsections that, based on joint cosmological constraints only,
overlap between surveys does not bring large advantages
compared to surveys covering disjoint regions of the sky.
One scenario in which cross-correlations between the two types of surveys
{\it can} be very advantageous is when the photometric redshift distribution
of the weak lensing survey has not been calibrated perfectly {\it a priori} \cite{newman08,mathnewman10,schulz10,mattnewman12,mcwhite13},
for example because the training sample of spectroscopic galaxies is insufficiently large or complete.
It has been shown in \cite{phzpaper} that in this case the cross-correlations between the number densities of
lensing source galaxies and spectroscopic galaxies (i.e.~the $ps$ spectra)
can help calibrate the photo-$z$ distribution and thus strongly improve the cosmology constraints
from weak lensing alone (in this case, the cosmological information in the spectroscopic galaxy
clustering was not used). Therefore, if we repeat the analysis of the previous sections, but this time
allowing for uncertainty in the photo-$z$ distribution, we might expect larger benefits from
having the two surveys overlap.

\begin{figure*}
  \begin{center}{
  \includegraphics[width=0.48\columnwidth]{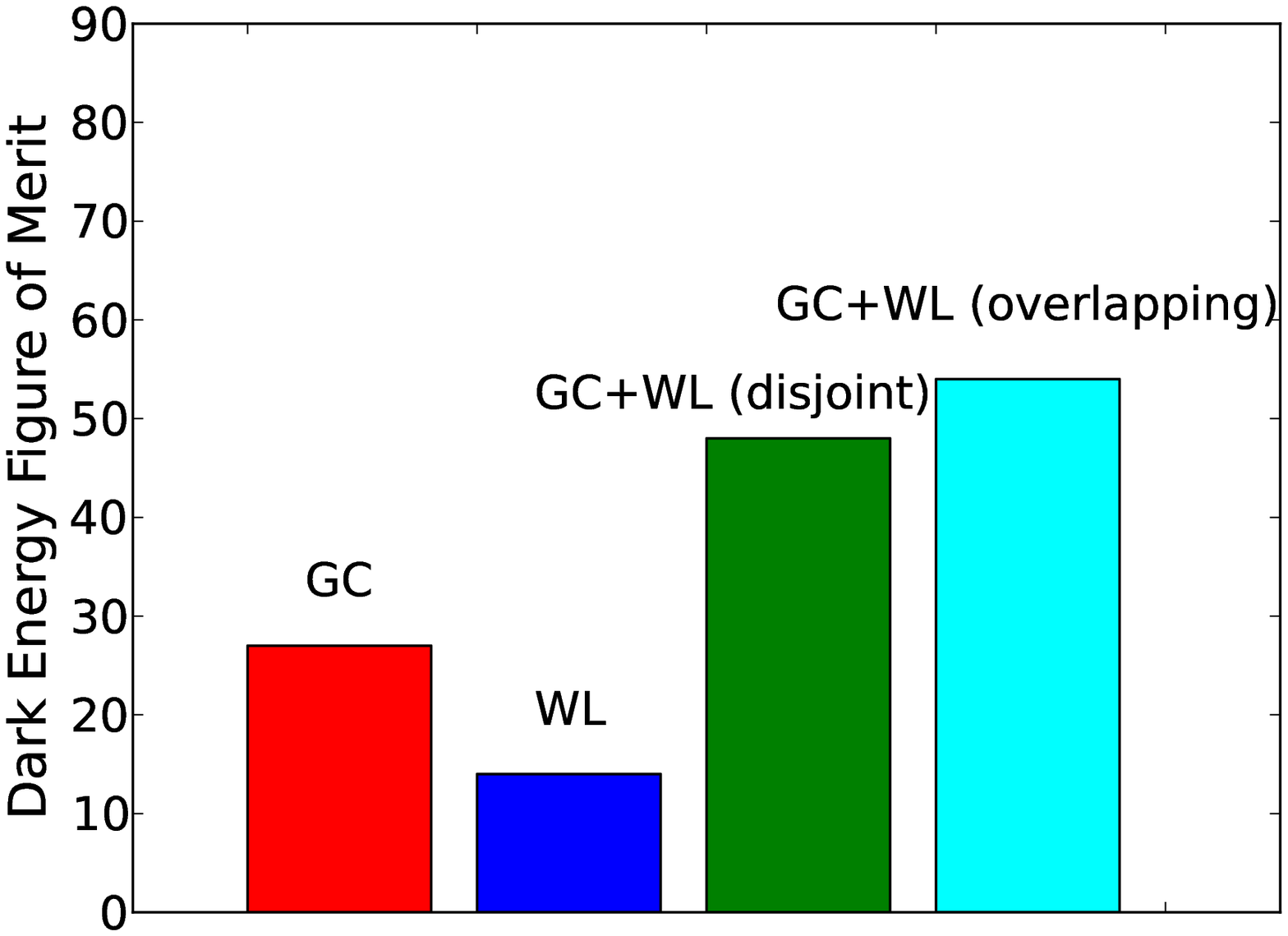} 
  \includegraphics[width=0.48\columnwidth]{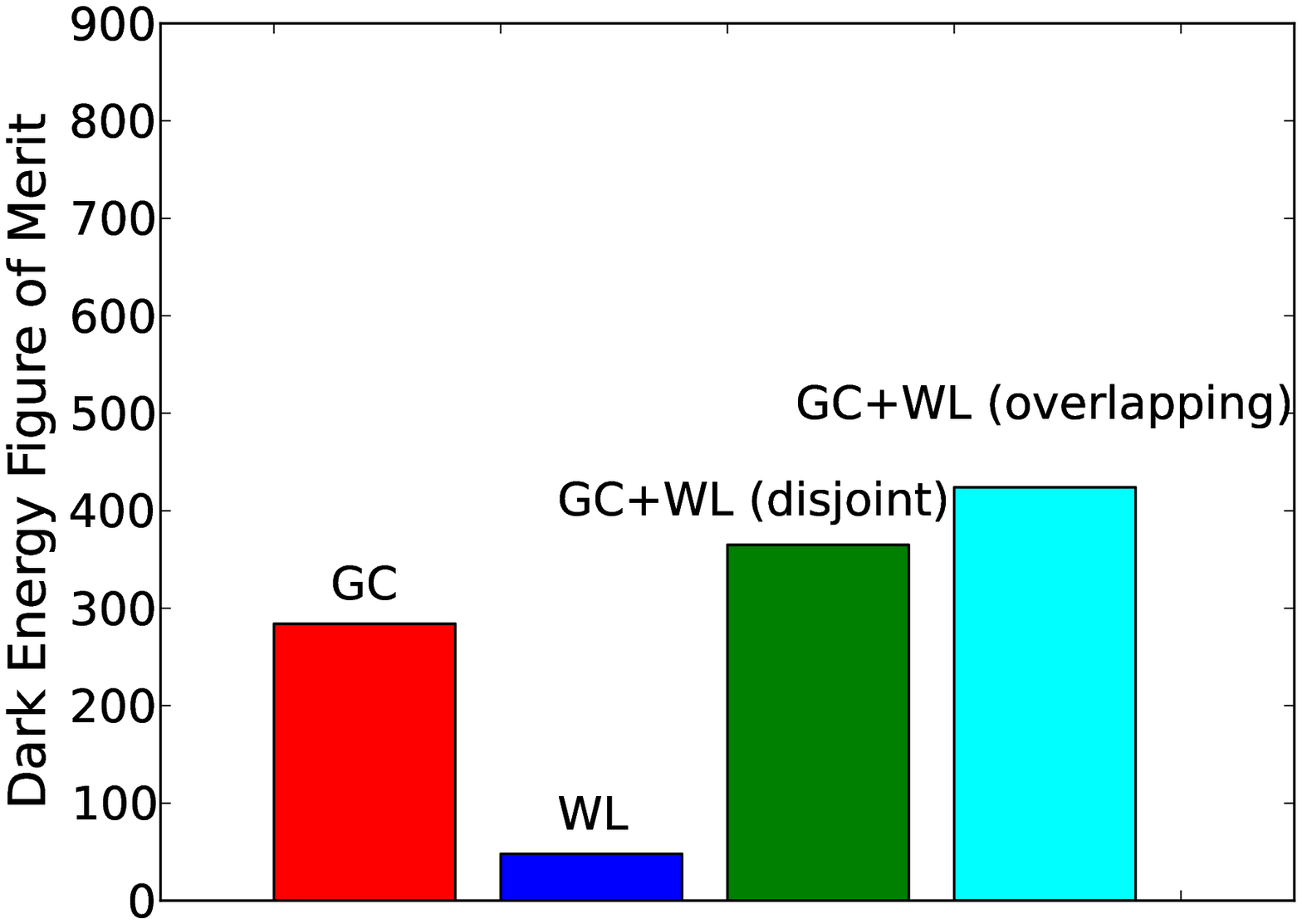} 
  }
  \end{center}
  \caption{Dark energy figures of merit from weak lensing and galaxy clustering when the photometric redshift distribution
  in the lensing survey is not known perfectly {\it a priori}. The distribution is modeled
  as a Gaussian characterized by a bias $b_z$ and a scatter $\sigma_z$.
  The redshift dependence of the bias and scatter is
  modeled as a spline with 11 nodes, giving 22 free photo-z parameters. The parameters
  have a prior $\sigma(\sigma_z) = \sigma(b_z) = 0.05$ and fiducial values such that
  $b_z(z) \equiv 0$ and $\sigma_z(z) = 0.05 (1 + z)$.
  Cross-correlations between spectroscopic and photometric galaxies help calibrate the photo-$z$
  distribution and thus slightly improve the benefits of overlapping surveys relative to
  disjoint surveys.
  {\it Left:} SuMIRe. {\it Right:} EUCLID.}
  \label{fig:phz fom}
\end{figure*}

To model the photometric redshift distribution, we follow the approach of \cite{phzpaper}
and treat the distribution $p(z_{\rm ph}|z)$ as a Gaussian. We thus ignore the possibility of outliers in the distribution.
For the bias and scatter, we assume,
as before, a fiducial $b_z(z) = 0$ and $\sigma_z(z) = 0.05 (1 + z)$. To model uncertainty in these quantities,
we describe both functions by a spline with 11 nodes evenly spaced in redshift in the range $z = 0 - 3$.
We assume the distribution is calibrated (e.g.~using a deep, matching spectroscopic sample)
at the level $\sigma(b_{z,i}) = \sigma(\sigma_{z,i}) = 0.05$, where $b_{z,i}$ and $\sigma_{z,i}$
are the values at the spline nodes. We then allow the data to self-calibrate the photo-$z$ parameters. As stated above,
specifically the $sp$ spectra are very useful for this. Unlike in \cite{phzpaper}, we still use the full cosmology information
present in the spectroscopic galaxy sample. The analysis is thus the same as in previous sections, except with the 22 photo-$z$
parameters added.

We show results for the dark energy figure of merit in Figure \ref{fig:phz fom}.
As expected, allowing for photo-$z$ distribution uncertainty
leaves the spectroscopic galaxy clustering constraints unchanged,
significantly weakens the FOM from weak lensing only, and weakens the joint constraints
to a lesser extent. The FOM is lowered less when surveys overlap than when they do not, because
overlapping surveys allow self-calibration using $sp$ cross-correlations. However, the difference is
still not spectacular: overlapping surveys give a $\sim 13 \%$ larger FOM for SuMIRe and a $\sim 16 \%$
larger FOM for EUCLID.
Finally, we note that we did not apply a prior to the bias of the lensing source galaxies here, and
that this could improve the photo-$z$ calibration somewhat and therefore the gains obtained from
overlapping surveys.

\section{why does overlap not matter?}
\label{sec:why}

We have shown so far for all surveys we considered that there is strong complementarity
between weak lensing (including clustering of source galaxies) and
spectroscopic galaxy clustering in the sense that combining the two leads to large improvements
in cosmological constraints. However, the role of overlap between surveys has proven to
be limited. This is somewhat surprising given some of the literature on this topic
\cite{caibern12,gaztaetal12,kirketal13}. To understand the lack of importance of overlap better,
let us consider in more detail the difference between the overlapping and disjoint survey scenarios.
In overlapping surveys, on the one hand, one can exploit the additional information present in the
cross-spectra (i.e.~$\gamma s$ and $p s$). On the other hand, one measures a smaller number of independent modes
when the surveys overlap (disjoint surveys offer twice the sky coverage). This loss of information
is also quantified by the cross-spectra. For instance, 
the covariance between the spectra $C_\ell^{ss}$ and $C_\ell^{\gamma \gamma}$ is given by
$2 (C_\ell^{\gamma s})^2/\left[ f_{\rm sky} (2 \ell + 1)\right]$.
One could thus imagine that our results are explained by the two above
mentioned effects cancelling out. However, we find that this is not the case, and that instead, both effects are small individually.
The real reason for overlap not making a difference is thus that the cross-spectra between probes contribute very little
compared to the auto-spectra.

The main reason for this is that the effective number of 3D density modes probed
by the cross-correlations is very small compared to the number of modes probed by either the
lensing auto-spectra or the galaxy clustering auto-spectra (see Fig.~5 of \cite{fontetal13}).
Heuristically, (spectroscopic) galaxy clustering probes a three-dimensional sphere in $k$-space
with radius given by $k_{\rm max} = 0.2 h/$Mpc, while shear probes mainly transverse modes in $k$-space,
but with a much larger non-linear cutoff, so that the numbers of modes in the two probes are within the
same order of magnitude. The cross-spectra, however, probe only the overlapping region
between the two volumes in $k$-space, i.e.~only transverse modes with a low cutoff $k_{\rm max} = 0.2 h/$Mpc.
This volume thus contains a much smaller number of modes. We make this argument more quantitative below.

Consider SuMIRe for example. Ignoring for now the lost information due to shot noise,
with our default choices of $k_{\rm max} = 0.2 h/$Mpc and (three tomographic bins with)
$\ell_{\rm max} = 2000$, the total number of spectroscopic galaxy and shear modes explicitly included are
$N^{\rm mode}_s = k_{\rm max}^3 V/(6 \pi^2) = 1.4 \times
10^6$ (where $V$ is the survey volume)
and $N^{\rm mode}_\gamma = f_{\rm sky} N_{\rm tom} \ell_{\rm max}^2 =
0.43 
\times 10^6$ respectively
(this is a rough estimate and the true number of available shear modes is determined by the redshift width of the shear
kernels rather than the number of tomographic bins).
The number of overlapping modes is less straightforward to quantify.
We cross-correlate $\sum_{j=1}^{N_s} \ell^2(k_{\rm max}, z_j) = 8.6 
\times 10^4$ transverse $s$-modes with the $\gamma-$ and $p-$modes
(with $\ell(k, z_j) \approx k D(z_j)$).
The number of shear (and also source density) modes is no larger than
$f_{\rm sky} \sum_{i=1}^{N_{\rm tom}} \ell^2(k_{\rm max}, z_i)$.
This gives $3.2 \times 10^4$ shear modes. Since this is smaller than the number of transverse $s$-modes,
the number of independent modes probed by cross-correlations cannot be larger than this,
$N^{\rm mode}_\perp < 3.2 \times 10^4$. This is merely $\sim 2 \%$ of $N^{\rm mode}_s$ and $\sim 7 \%$ of
$N^{\rm mode}_\gamma$.

\begin{figure*}
  \begin{center}{
  \includegraphics[width=0.48\columnwidth]{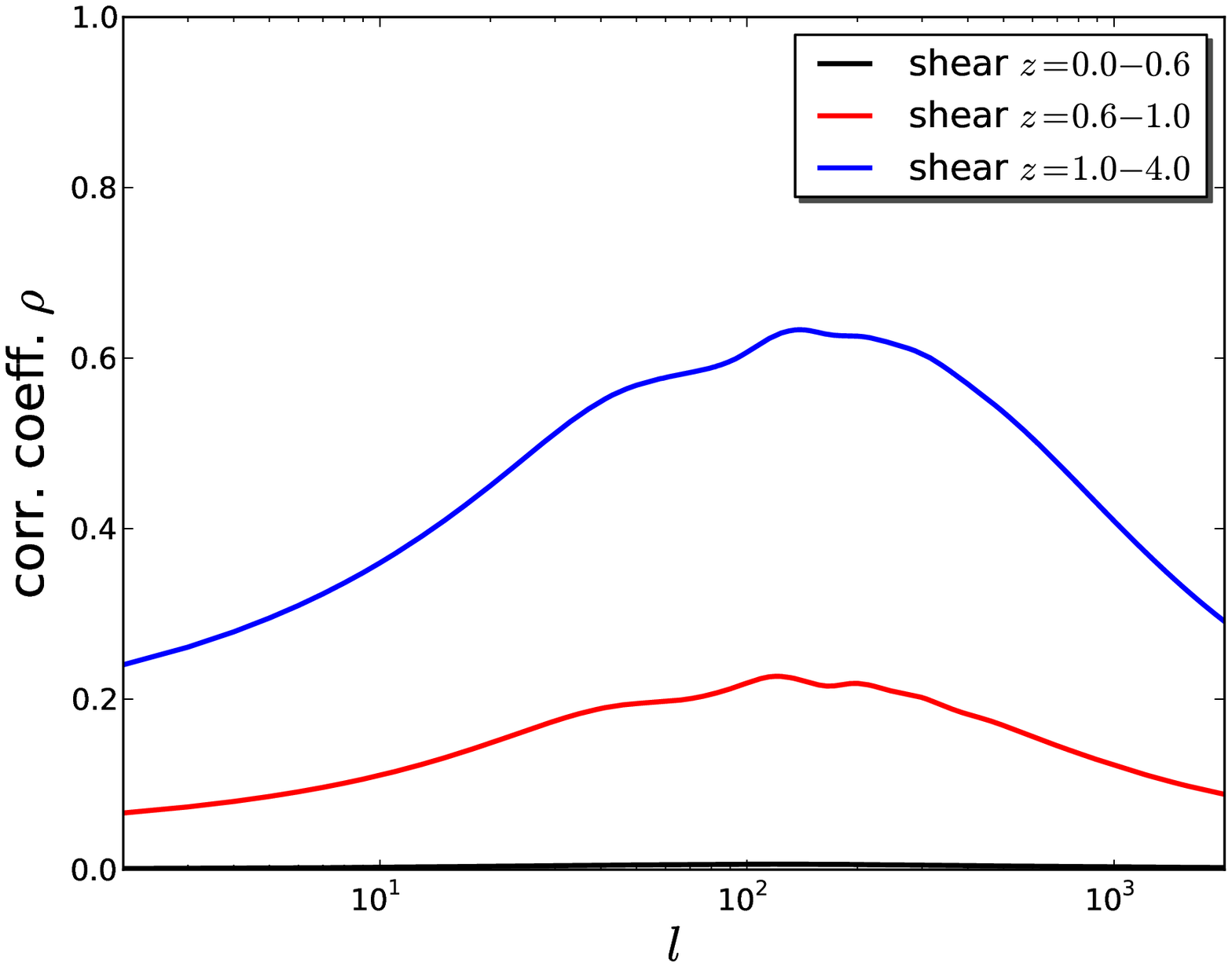}
  \includegraphics[width=0.48\columnwidth]{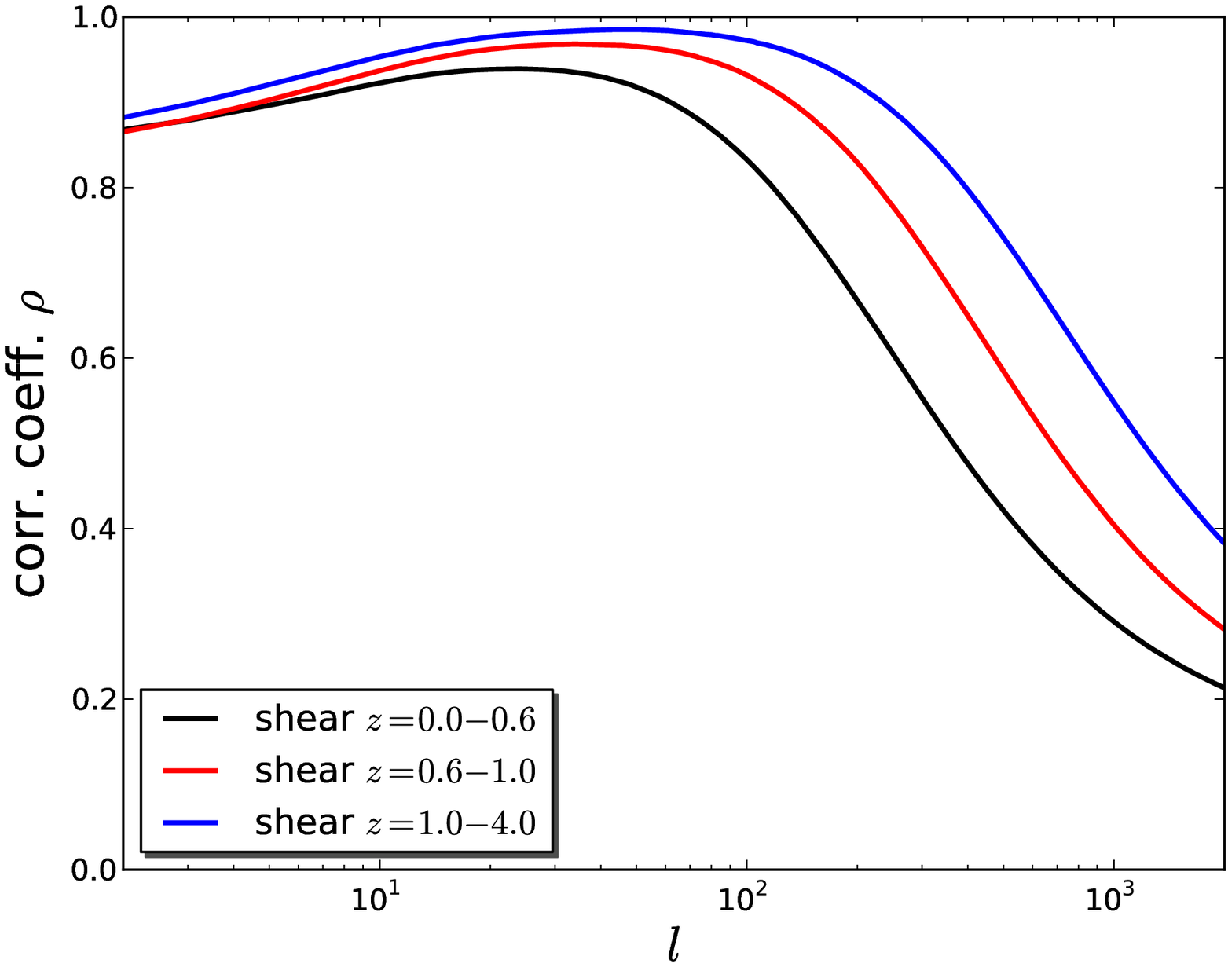}
  }
  \end{center}
  \caption{The correlation coefficient $\rho$ of a shear mode in a given tomographic bins with {\it the set of}
  spectroscopic modes in all redshift bins, as a function of $\ell$. {\it Left:} SuMIRe. The correlation of the lensing
  modes with the spectroscopic galaxies is far from 100\% because the overlap between the spectroscopic galaxy
  redshift distribution and the lensing kernels is limited (the
  spectroscopic sample lacks coverage at $z < 0.6$). Note that the $z_{\rm ph} = 0 - 0.6$ lensing source bin does have some galaxies at
  $z>0.6$ due to the photo-z
	scatter, explaining why the black curve is not exactly equal to zero.
  {\it Right:}  When galaxies at $z = 0 - 0.6$ are added to the spectroscopic survey the correlation becomes
  close to maximal, except at high multipoles.
  At a given multipole $\ell$, the correlation coefficient for tomographic bins $i$ is defined as $\rho \equiv \sqrt{1 - ({\bf F}_{\gamma_i \gamma_i} {\bf C}_{\gamma_i \gamma_i})^{-1}}$,
  where ${\bf C}$ is the covariance matrix of the modes $\gamma_i$ and $\{s_j\}_{j=1}^{N_s}$ and ${\bf F}$ is
  its inverse (we suppress the $\ell$-dependence of the previous
 expressions here). In other words, $\rho$ is the generalization of
 $r=C^{\gamma s}_l/
\sqrt{C^{\gamma\gamma}_lC^{ss}_l}$
  to the case of multiple $s$ fields.
}
  \label{fig:corrcoeff}
\end{figure*}

On top of this, it needs to be taken into account that the correlation between the shear modes and the spectroscopic modes
is not optimal. Figure \ref{fig:corrcoeff} (left panel)
shows the effective correlation coefficient $\rho$ of each shear mode, as a function of multipole,
with the {\it full set of} transverse spectroscopic galaxy density modes. We do not include shot noise or any non-linear cutoff
in this plot. For a given shear mode, if the set of $s$-modes probed all the 3D density modes contributing to that shear mode,
the correlation would be optimal, $\rho = 1$. Instead, we see that $\rho$ is significantly smaller than unity.
The main reason for this is the lack of overlap in redshift between the shear kernel and the spectroscopic galaxy
distribution. The mean source redshift is $z \sim 1$ so that the typical shear kernel peaks halfway between $z = 0$
and $z = 1$. However, SuMIRe only includes galaxies at $z > 0.6$, thus missing a large fraction of the lensing kernel.
The right panel of figure \ref{fig:corrcoeff} shows how the correlation coefficient increases when spectroscopic galaxies are added at
$z = 0 - 0.6$.
This is in fact a very realistic scenario to consider, as spectroscopic galaxy clustering information
from SDSS and BOSS can be included at $z < 0.6$.
Now the correlation is much stronger although still not identically equal to one.
However, while imposing full redshift overlap (by adding galaxies at $z=0-0.6$) strengthens the cross-correlation signal,
we have checked that even after the $z = 0 - 0.6$ galaxies have been added, the
effect of survey overlap on dark energy and modified gravity constraints is negligible, thus confirming
that the dominant reason for the lack of complementarity is the number
of modes probed by the cross-correlations.

The counting of modes presented above does not include the effect of shot noise, which strongly reduces
the effective number of modes accessible at small angular scales. As an
alternative proxy
for the effective number of
modes we therefore consider
\beq
N_{\rm eff} \equiv 2 (S/N)^2 = 2 \sigma_8^2 F_{\sigma_8, \sigma_8},
\eeq
where $F_{\sigma_8, \sigma_8}$ is the diagonal Fisher matrix element corresponding to the amplitude $\sigma_8$.
Thus, $S/N$ is simply the signal to noise of detecting the amplitude of the signal, including all modes up to the cutoff.
In the case of a single angular power spectrum (i.e.~a single redshift bin), this becomes
\beq
N_{\rm eff} = 2 \sum_{\ell} \frac{C_l^2}{\sigma^2(C_l)}
\eeq
(the expression for a 3D power spectrum is very similar).
If $C_l$ is an auto-power spectrum (as opposed to a cross-spectrum), in the absence of shot-noise,
we have
$\sigma^2(C_l) = 2 (C_l)^2/\left[ f_{\rm sky} (2 \ell + 1)\right]$, so that
\beq
N_{\rm eff} = \sum_{\ell} f_{\rm sky} (2 \ell + 1),
\eeq
which is exactly the total number of modes. The presence of shot noise
increases the variance $\sigma^2(C_l) = 2 (C_l + N_l)^2/\left[ f_{\rm sky} (2 \ell + 1)\right]$
and thus reduces $N_{\rm eff}$. In the case of cross-correlations, $N_{\rm eff}$ also takes into account
the reduced information due to the two fields not being optimally correlated.
In the case of no shot noise, we for example have for the effective number of
modes probed by $\gamma s$,
\beq
N_{\rm eff} = \sum_{\ell} \frac{r^2}{1 + r^2} (2 \ell + 1) f_{\rm sky}, \quad \quad r \equiv \frac{C_\ell^{\gamma s}}{\sqrt{C_\ell^{\gamma \gamma} C_\ell^{ss}}}.
\eeq
A correlation coefficient $r < 1$ causes a reduction in the effective number of modes.

Using the above definition of $N_{\rm eff}$, we find
$N_{\rm eff} = 8.3 \times 10^5$ for the spectroscopic survey alone,
$N_{\rm eff} = 4.4 \times 10^5$ for the imaging survey (combining the $\gamma$ and $p$ modes),
while the cross-correlations (i.e.~$\gamma s$ and $ps$) gives
$N_{\rm eff} = 4.4 \times 10^4$, which is a factor ten smaller than the number of modes probed by lensing
and a factor twenty below the number probed by galaxy clustering.

In conclusion, given the limited effective number of modes available in the cross-spectra,
it is not surprising that the overlap between surveys does not significantly affect the forecasted
cosmology constraints. We find qualitatively similar mode counting results for the other surveys.

%



\section{Conclusions and Discussion}
\label{sec:conclusions}

\subsection{Summary of Results}


We have studied dark energy and modified gravity constraints that can be obtained from the combination
of weak lensing and spectroscopic galaxy redshift data that will be available
from the SuMIRe, EUCLID and WFIRST surveys. For the weak lensing components of these surveys,
we considered galaxy shear in tomographic bins, and the overdensity of lensing source galaxies.
We assumed the lensing source galaxies have tomographic redshifts. For the spectroscopic galaxy clustering
components of the surveys, we considered the three-dimensional, redshift-space overdensity field of the galaxies.
Using the Fisher matrix formalism, we have quantified the information encoded in all available two-point functions
of these observables, on large (linear and quasi-linear) scales. We always included a prior from
Planck CMB data and marginalized over cosmological (and galaxy bias) parameters, including neutrino mass.

In all cases, we have found strong complementarity between the two probes. The dark energy figure of merit is up to a factor
of $\sim 2.5$ larger when the probes are combined than for the strongest of the individual probes, and all three surveys
promise strong constraints on the dark energy equation of state. Even more dramatically,
treating the growth parameter, $f(z)$, as a free function (i.e.~allowing for deviations from general relativity),
we find that the combination of a weak lensing with a galaxy clustering survey can constrain
$f(z)$ at the few percent level, while each survey individually has negligible constraining power.

While all three combinations of surveys studied here have full sky overlap of the weak lensing and spectroscopic components,
we have also studied how the forecasted constraints depend on survey overlap. We have done this by comparing forecasts for
the fully overlapping case to the case where the surveys are described by the exact same specifications, except
that the imaging and spectroscopic components now cover disjoint regions of the sky. We have found for all three survey combinations
that the difference in constraining power between these two scenarios is small (typically $< 10 \%$ differences in
uncertainties and figures of merit). Following \cite{fontetal13} (see their Appendix B for a clear, qualitative explanation),
we attribute this to the small number of three-dimensional density modes probed by the cross-correlations between the survey
observables, as compared to the number of modes probed by the auto-correlations within each survey. We have shown,
that the number of modes probed by the cross-correlations depends on the method used to do the calculation, but is
no more than $10 \%$ of the smallest number of modes probed by each survey individually. Moreover, we
found that for the surveys under consideration in this work, the limited redshift overlap between the spectroscopic redshift distribution
and lensing kernels
weakens the cross-correlations between these observables. However, while enhancing the redshift coverage of
the spectroscopic survey by adding galaxies at low redshift improves the level of correlation between the surveys, we have checked that the same-sky benefit
is small even in that case.

Finally, we have included the possibility of uncertainty in the parameters describing the photo-$z$
distributions, following the methodology described in \cite{phzpaper}. In this case, it is known that cross-correlations of the lensing source galaxies with a spectroscopic sample
can help calibrate the photo-$z$ distribution. Thus, one might expect larger same-sky benefits in this case.
We found, however, that this is not a large effect. The boost in dark energy figure of merit due to
covering the same sky area is at most $\sim 20 \%$, when photo-$z$ scatter and bias are treated as free parameters.

\subsection{Comparison with Literature}

It is worth commenting on the current status in the literature of the question
of same-sky benefits, i.e.~the question of how much better (if at all) dark energy and/or modified gravity
constraints from overlapping survey are compared to those from disjoint surveys.
There are a number of groups that have recently addressed this issue, or are in the process
of doing so, but results vary strongly. To crudely summarize,
\cite{caibern12} find only modest same-sky benefits for realistic survey galaxy number densities
and \cite{fontetal13} also finds results consistent with ours. On the other hand,
\cite{gaztaetal12} found enormous increases in figure of merit (up to a factor of 100, although not the same figure of merit as
considered here), and, more recently, \cite{kirketal13} also found large same-sky improvement factors (a factor $\sim 4$ for dark energy
in their abstract). In addition, several groups \cite{songinprep,gaztainprep} are working on new results,
and thus add to the variety of answers.

While a comparison between the various groups is not easy because of different choices of survey properties,
it is unlikely that survey specifications can explain why some groups find large ($\gtrsim 4$) improvement
factors, while others find only modest $\sim 1 - 1.5$ improvement factors, if any.
An interesting work to compare with is the one by \cite{kirketal13}, first of all, because it is public
(available on the arXiv) and, secondly, because it presents such different results compared to ours.
The major difference between \cite{kirketal13} on the one hand and the present article (and also \cite{caibern12,gaztaetal12,fontetal13})
on the other hand, is the forecast method. While we consider the three-dimensional power spectrum of spectroscopic galaxies,
\cite{kirketal13} treat the spectroscopic galaxy density purely in terms of angular power and cross-spectra.
While a large number of spectroscopic redshift bins is used, $N_s = 40$, this still implies
a bin width $\Delta z = 0.0425$, corresponding to a comoving distance $\sim  130 - 50 h^{-1}$Mpc (at $z = 0 - 1.7$),
over which line-of-sight clustering information is lost. On the upside, this approach allows the use of a single, consistent method
to describe all data and their covariances, and no Limber approximation is used to compute the angular spectra.

We have tried to reproduce the same-sky benefits of \cite{kirketal13} by emulating their survey assumptions
as much as possible given the information provided in the article. Moreover, to mimic their treatment of the galaxy clustering
information, we have performed calculations either using angular spectra in 40 bins (however, still using the Limber approximation)
or by using a three-dimensional power spectrum, but with a line-of-sight degradation factor smoothing out information
on scales smaller than that of the bin width. While neither of these methods matches exactly the approach of
\cite{kirketal13} (because we are not set up to do a forecast based on exact angular power spectra, i.e.~without assuming
the Limber approximation), one would expect to be able to at least approximately reproduce their results.
Unfortunately,
even with these changes, we cannot reach improvement factors for the dark energy figure of merit better than $\sim 1.4$,
while \cite{kirketal13} find factors of $3 - 5$ for their case of scale-independent galaxy bias (although we are not sure whether
this result includes the CMB prior or not).

We have also checked that the lack of same-sky benefits in our forecasts
is not driven by our perhaps optimistic assumptions about modeling of galaxy clustering
in the non-linear regime, $k_{\rm max} = 0.2 h/$Mpc (although, we remind the reader that we always
exponentially suppress information at scales $\ls 10 h^{-1}$Mpc to take into account bulk flows \cite{seoeis07}).
To this end, we have repeated our analysis for SuMIRe with $k_{\rm max} = 0.1 h/$Mpc.
While the resulting dark energy and growth constraints are significantly weaker
than for $k_{\rm max} = 0.2 h/$Mpc, the results remain the same qualitatively:
combining the two surveys strongly improves constraints, but it hardly matters whether the surveys
overlap or not.

While we are very confident in our results, it is important to resolve/explain the differences between groups
and for the community to converge on a single answer to and understanding of the question of same-sky benefits,
as this is an important consideration when designing future surveys.
There has been some effort to resolve the differences as part of a MS-DESI
working group, but clearly more work needs to be done to reach a resolution.

\subsection{Other Sources of Synergy}

Although throughout this
paper we have focused on the large-scale galaxy clustering information up to $k_{\rm
max}=0.2~h/{\rm Mpc}$ to avoid uncertainties inherent in non-linear
processes such as galaxy formation, several groups
\cite{Cacciatoetal:13,Reddicketal:13,hikageetal12,hikageetal12b,HikageYamamoto:13}
have proposed promising methods for using the small-scale clustering
signal, where the signal-to-noise ratio is much higher, in order to
infer the connection of galaxies to dark matter halos. In particular,
\cite{Cacciatoetal:13,hikageetal12b} showed that the cross-correlations
of spectroscopic galaxies with shapes of background galaxies, or with the
positions of photometric galaxies at similar redshifts to the
spectroscopic galaxies, are useful to constrain the galaxy-halo connection
and then improve the cosmological interpretation of the large-scale
clustering signal. The small-scale clustering information might thus
offer a further promising synergy of imaging and spectroscopic galaxies,
beyond what we discussed in this paper, if our understanding of galaxy
formation at small scales is improved, or a robust, empirical method to calibrate these
uncertainties is developed.
These methods can only be applied if the imaging and spectroscopic
surveys overlap on the sky.

Other types of synergy between overlapping surveys not studied in this work
are the fact that the imaging survey can be used to create a target catalog
for the spectroscopic survey, and the fact that the combination of the
two data sets will be robust against systematics that affect only weak lensing,
or only galaxy clustering.

\section{acknowledgments}

Part of the research described in this paper was carried out at the Jet Propulsion Laboratory, California Institute of Technology,
under a contract with the National Aeronautics and Space
Administration. This work is supported by NASA ATP grant 11-ATP-090.
MT was supported by World
 Premier International Research Center Initiative (WPI Initiative),
 MEXT, Japan, by the FIRST program ``Subaru Measurements of Images and
 Redshifts (SuMIRe)'', CSTP, Japan, and by Grant-in-Aid for
 Scientific Research from the JSPS Promotion of Science (23340061).
We also acknowledge the input of Sudeep Das, who wrote an early version of the Fisher matrix code which some of
the code we used for this work builds on.

\bibliography{refs}

\end{document}